\documentclass[12pt]{article}
\usepackage{epsf,latexsym}
\usepackage{amsfonts,amssymb} 
\epsfverbosetrue
\newcommand{\double}[1]{\mathbb{#1}}

\newcommand{\cc}{\double{C}}

\newcommand{\rr}{\double{R}}
\newcommand{\zz}{\double{Z}}

\newcommand{\aaa}{\mathcal{A}}
\newcommand{\ccc}{\mathcal{C}}

\newcommand{\hhh}{\double{H}}
\newcommand{\mm}{\mathcal{M}}

\newcommand{\pp}{\pmatrix}

\newcommand{\dd}{\mathcal{D}}

\newcommand{\hh}{\mathcal{H}}
\newcommand{\ff}{\mathcal{F}}
\newcommand{\llll}{\mathcal{L}}
\newcommand{\sss}{\mathcal{S}}
\newcommand{\jj}{\mathcal{J}}

\newcommand{\ttt}{{\rm tr}}

\def\ddd{{\,\hbox{$\partial\!\!\!/$}}}

\newcommand{\dee}{\hbox{\rm{D}}}
\newcommand{\de}{\hbox{\rm{d}}}

\newcommand{\pa}{\partial}

\newcommand{\ot}{\otimes}
\newcommand{\op}{\oplus}

\newcommand{\bb}{\begin{eqnarray}}
\newcommand{\ee}{\end{eqnarray}}
\newcommand{\eee}{\nonumber\end{eqnarray}}
\newcommand{\qq}{\quad}
 
\begin{document}

\font\twelve=cmbx10 at 13pt
\font\eightrm=cmr8

\thispagestyle{empty}

\begin{center}

CENTRE DE PHYSIQUE TH\'EORIQUE $^1$ \\ CNRS--Luminy, Case 907\\ 13288 Marseille Cedex 9\\
FRANCE\\

\vspace{2cm}

{\Large\textbf{ Noncommutative Geometry and\\[2mm]
the Standard Model}} \\

\vspace{1.5cm}

{\large Thomas Sch\"ucker $^2$}

\vspace{1.5cm}

{\large\textbf{Abstract}}
\end{center}

Connes' noncommutative approach to the standard model of
electromagnetic, weak and strong forces is sketched as well as its unification with
general relativity.
\vspace{1.5cm}

\vskip 1truecm

PACS-92: 11.15 Gauge field theories\\
\indent MSC-91: 81T13 Yang-Mills and other gauge theories

\vskip 1truecm

\noindent 
\vskip 1truecm
\noindent CPT-2004/P.049\\

\vspace{2cm}
\noindent $^1$ Unit\'e Mixte de Recherche 
(UMR 6207)
 du CNRS  et des Universit\'es Aix--Marseille 1 et 2 et  Sud
Toulon--Var, Laboratoire affili\'e \`a la FRUMAM (FR 2291)\\
$^2$ Also at Universit\'e Aix--Marseille 1, \\
 schucker@cpt.univ-mrs.fr 

\section{Spectral triples}

Noncommutative geometry \cite{book,real,grav} equips Riemannian spin manifolds
with an uncertainty relation just as quantum mechanics equips phase space with
Heisenberg's uncertainty relation. Major building blocks of the theory are therefore
the algebra of observables and a representation on a Hilbert space. More precisely a
{\bf real, even spectral triple } is given by five items
\begin{itemize}\item
$\aaa$ is a real, associative algebra with unit 1 and involution $*$. (Its elements can
be called observables.)
\item
$\hh$ is a complex Hilbert space carrying a faithful representation $\rho $ of the
algebra. (Its elements are the wave functions.)
\item
$\dd$ is a selfadjoint operator on $\hh$ of compact resolvent. We call it a Dirac
operator.
\item
$J$ is an anti--unitary operator on $\hh$. We call it real structure or charge
conjugation.
\item
$\chi $ is a unitary operator on $\hh$. We call it chirality.
\end{itemize} We require the following axioms to hold:
\begin{itemize}\item
$J^2=-1$ in four dimensions (\,$J^2=1$ in zero dimension).
\item
$[\rho(a),J\rho(\tilde a)J^{-1}]=0$ for all $a,\tilde a\in\aaa$.
\item
$\dd J=J\dd,\qq J\chi=\chi J,\qq \dd\chi=-\chi\dd$. 
\item
$[\dd,\rho(a)] $ is bounded for all $a\in\aaa$ and $[[\dd,\rho(a)],J\rho(\tilde
a)J^{-1}]=0$ for all $a,\tilde a\in\aaa$. This property is called first order condition
because in the first  example below, it states that the genuine Dirac operator is a first
order differential operator.
\item
$\chi^2=1$ and $[\chi ,\rho (a)]=0$ for all $a\in\aaa$. These properties allow the
decomposition $\hh=\hh_L\op\hh_R$.
\item We suppose the kernel of $\dd$ to contain no non--trivial subspace invariant
under the representation of the algebra $\aaa$ (``non--degeneracy'').
\item
   There are three more properties, that we do not spell out, orientability, which
relates the chirality to a volume form, Poincar\'e duality and regularity, which
states that the observables $a\in\aaa$ are differentiable.
\end{itemize}

The calibrating, commutative example is given by an even dimensional, compact, 
Riemannian spin manifold $M$ (Euclidean signature), which for concreteness we
take to be 4-dimensional (Euclidean timespace). Let
$\aaa=\ccc^\infty(M)$, the algebra of complex valued differentiable functions,
$\hh=\llll^2(\sss)$, the square integrable sections of the spinor bundle,
$\dd=\ddd$, the genuine Dirac operator, $J=\gamma ^0\gamma ^2 \circ$ complex
conjugation and $\chi = \gamma _5=\gamma ^0\gamma ^1\gamma ^2\gamma ^3$.
The $\gamma ^a$ are the Dirac matrices.

Another commutative example, the two--point space, is discrete or 0-dimensional:
\bb\aaa= \cc_L\op\cc_R\owns (a_L,a_R),\qq
\hh=\cc^4,\qq
\rho(a_L,a_R)=\pp{a_L&0&0&0\cr 0&a_R&0&0\cr 0&0&\bar a_R&0\cr 0&0&0&\bar
a_R},\ee
\bb\dd =\pp{0&m&0&0\cr \bar m &0&0&0\cr 0&0&0&\bar m\cr0&0&m&0 },\qq
 J=\pp{0&1_2\cr 1_2&0}\circ {\rm c\ c},\qq
\chi =\pp{-1&0&0&0\cr 0&1&0&0\cr 0&0&-1&0\cr 0&0&0&1}.\ee  As we will see, the
parameter $\ m\in\cc$ will determine the distance between the two points as
$1/|m|$.  For 0-dimensional triples the regularity axiom is empty, while orientability
means that the
 chirality can be written as a finite sum,
\bb \chi = \sum_j \rho (a_j)J\rho (\tilde a_j)J^{-1},\qq a_j,\tilde a_j\in\aaa.\ee The
Poincar\'e duality says that the intersection form  
\bb \cap_{ij}:=\ttt
\left[\chi\,\rho(p_i)\,J\rho(p_j)J^{-1}
\right]\ee
 must be non-degenerate, where the $p_j$ are a set of minimal projectors of $\aaa$.
The two-point space is orientable, $\chi =\rho (-1,1)J\rho (-1,1)J^{-1}$. It also
satisfies Poincar\'e duality, there are two minimal projectors, $p_1=(1,0)$,
$p_2=(0,1)$, and the intersection form is
$\cap=\pp{0&-1\cr -1&2}$.

Connes' reconstruction theorem \cite{grav} states that all {\it commutative,} real
(even) spectral triples are of the above type: they come from a (even or
0-dimensional) compact Riemannian spin manifold $M$. In this reconstruction the
Dirac operator plays three roles: it allows to reconstruct the dimension, the metric
and the differential structure:

The {\bf  dimension} is a local property of space. It can be retrieved from the
asymptotic behaviour of the spectrum of the Dirac operator for large eigenvalues.
Since
$M$ is compact, the spectrum is discrete. Let us order the eigenvalues, $...\lambda
_{n-1}\leq\lambda _n\leq\lambda _{n+1}...$ Then Weyl's spectral theorem states that
the eigenvalues grow asymptotically as
$n^{1/{\rm dim}M}$. To explore a local property of space we only need the high
energy part of the spectrum. This is in nice agreement with our intuition from
quantum mechanics and motivates the name `spectral triple'.

The {\bf metric} can be reconstructed from the commutative spectral triple by
Connes' distance formula (\ref{dist}) below. In the commutative case points
$x\in M$ are reconstructed as pure states. The general definition of a pure state of
course does not use the commutativity. A state $\delta  $ of the algebra $\aaa$ is a
linear form on $\aaa$, that is normalized, $\delta (1)=1$, and positive, $\delta 
(a^*a)\geq 0$ for all $a\in\aaa$. A state is pure if it cannot be written as a convex
combination of two states. For the calibrating example, there is a one-to-one
correspondence between points $x\in M$ and pure states $\delta  _x$
 defined by the Dirac distribution,
$\delta  _x(a):=a(x)=\int_M\delta _x(y) a(y)\de^4 y$. The geodesic distance between
two points $x$ and $y$ is reconstructed from the triple as:
\bb {\rm sup}\left\{ |\delta _x(a)-\delta _y(a)|;\ a\in\ccc^\infty(M)\ {\rm such \
that}\ ||[\ddd,\rho (a)]||\leq 1\right\} .\label{dist}\ee Note that Connes' distance
formula continues to make sense for non-connected manifolds, like discrete spaces.

{\bf Differential forms,} for example of degree one like
$\de a$ for a function $a\in\aaa$, are reconstructed as
$(-i)[\ddd,\rho (a)]$. This is again motivated from quantum mechanics. Indeed in a
1+0 dimensional timespace $\de a$ is just the time derivative of the observable $a$
and is associated with the commutator of the Hamilton operator with $a$.

As in quantum mechanics, we define a noncommutative geometry by a real spectral
triple with noncommutative algebra
$\aaa$. Here is a 0-dimensional noncommutative example: 
\bb \aaa=\hhh\op\cc\op M_3(\cc)\,\owns\,(a,b,c), \qq \hh=\cc
^{30}.\label{ncex6}\ee  By $\hhh$ we denote the quaternion algebra, complex
$2\times 2$ matrices of the form
\bb a=\pp{x&-\bar y\cr y&\bar x}.\ {\rm The\  representation\  is:}\ \rho(a,b,c):=
\pp{\rho_{L}&0&0&0\cr 0&\rho_{R}&0&0\cr 0&0&{\bar\rho^c_{L}}&0\cr
0&0&0&{\bar\rho^c_{R}}}\ee with
\bb\rho_{L}(a):=\pp{ a\ot 1_3&0\cr 0&a&},\qq
\rho_{R}(b):= \pp{ b  1_3&0&0\cr 0&\bar b  1_3&0\cr 0&0&\bar b}, \label{repr1}
\ee\bb
  \rho^c_{L}(b,c):=\pp{ 1_2\ot c&0\cr 0&\bar b1_2},\qq
\rho^c_{R}(b,c) := \pp{
 c&0&0\cr 0& c&0\cr 0&0&\bar b},  \label{repr2}
\ee
\bb \dd=\pp{0&\mm&0&0\cr
\mm^*&0&0&0\cr 0&0&0&\bar\mm\cr 0&0&\bar\mm^*&0},\qq
\mm=\pp{
\pp{m_u&0\cr 0&m_d}\ot 1_3 &0\cr 0&\pp{0\cr m_e}},\label{mass}\ee with three
positive numbers $m_u,\  m_d$ and $m_e$,
\bb J=\pp{0&1_{15}\cr 1_{15}&0}\circ
\ {\rm complex\ conjugation},\qq
 \chi=\pp{-1_{8}&0&0&0\cr 0&1_{7}&0&0\cr
 0&0&-1_{8}&0\cr 0&0&0&1_{7} }.\label{ncex11}\ee This triple is orientable, $\chi =
\rho (-1_2,1,1_3)J\rho (-1_2,1,1_3)J^{-1},$ and Poincar\'e duality holds. We have
 three minimal projectors,
\bb p_1=(1_2,0,0),\qq p_2=(0,1,0), \qq p_3=
\left(0,0,\pp{1&0&0\cr 0&0&0\cr 0&0&0}\right)\ee and the intersection form
\bb \cap=-2\pp{0&1&1\cr 1&-1&-1\cr 1&-1&0},\ee is non-degenerate.

One of the attractive features of spectral triples is that they describe continuous and
discrete spaces in the same language and therefore define differential forms also for
discrete spaces. Another attractive feature is tensorisation. Given two real, even
spectral triples, $(\aaa_j,
\hh_j,...)$, $j=1,2$, we get a third by taking $\aaa=\aaa_1\ot\aaa_2$,
$\hh=\hh_1\ot\hh_2$,
$\dd=\dd_1\ot 1_2 + \chi _1\ot \dd_2$, $J=J_1\ot J_2$ and $\chi =\chi  _1\ot\chi _2$.
The other
 obvious choice for the Dirac operator,
$\dd_1\ot\chi _2\ +\ 1_1\ot\dd_2$, is unitarily equivalent to the first one. In the
commutative case this tensor product simply describes the direct product of spaces,
$\ccc^\infty(M_1)\ot\ccc^\infty(M_2)=\ccc^\infty(M_1 \times M_2)$, 
$\llll^2(\sss(M_1))\ot\llll^2(\sss(M_2))=\llll^2(\sss(M_1\times M_2))$,...

We define a (4-dimensional) {\bf almost commutative  spectral triple} to be  the
tensor product of the commutative spectral triple coming from a (4-dimensional)
space $M$ and a noncommutative 0-dimensional spectral triple. An almost
commutative space therefore has an infinite number of commutative degrees of
freedom plus a finite number of noncommutative degrees of freedom. Motivation for
almost commutative geometries comes from Pauli spinors where the Pauli matrices
generate the quaternions or from Kaluza-Klein theories with the fifth dimension
discrete and fuzzy \cite{madore}. Note that the genuine Dirac operator on $M$ is
massless. By the tensorisation with  0-dimensional triples, commutative or not, it
becomes massive and $\mm$ is the fermionic mass matrix.

At present, we only have a few examples of truly noncommutative geometries, the
first being the kinematics of quantum mechanics:  in its compact version the
noncommutative torus \cite{nct} or in its noncompact version the Moyal plane
\cite{moyal}.

\section{The spectral action}

Every spectral triple carries a natural action, the spectral action \cite{grav,cc}. Its
configuration space $\ff$ is an affine space constructed by extending Einstein's
equivalence principle to noncommutative spaces: 
\bb\ff:=\left\{\dd_f= \sum_{\rm finite} r_jL(\sigma _j) \dd L(\sigma _j)^{-1}, \ 
r_j\in\rr,\
\sigma _j\in {\rm Aut}(\aaa), \right\}.\ee Indeed in the commutative case, every
automorphism of $\ccc^\infty(M)$ 
 is a diffeomorphism $\sigma $ of
$M$, a general coordinate transformation. $L(\sigma )$ is its double--valued
lift \cite{lift} acting on spinors
$\psi \in \llll^2(\sss)$, in a coordinate neighbourhood
\bb \big( L(\sigma  )\psi \big) ( x)=\big(S\left(\Lambda (\sigma,g
)\big)\right|_{\sigma  ^{-1}( x)}\psi
\big({\sigma ^{-1}( x)}\big),\ee  with
\bb 
\begin{array}{clc} S: SO(n) &\longrightarrow  \, Spin(n)& 
\left.\Lambda(\sigma ,g)\right|_x =\left[\sqrt{\jj_\sigma ^{-1T}\,g\,\jj_\sigma
^{-1}}\,\jj_\sigma\, 
\sqrt{g^{-1}}\right]_x, \\
\Lambda =\exp\omega  &\longmapsto \, \exp \left({\textstyle\frac{1}{8}} 
\omega_{ab} [\gamma ^{a},\gamma ^b]\right),&{\jj_\sigma ( x) ^{\nu 
}}_{\mu}:={\pa\sigma
 ^{\nu }(x) }/{\pa x^\mu} ,
\end{array}\label{spin}\ee  
and $g_{\mu \nu }:=g(\pa /\pa x^\mu , \pa /\pa
x^\nu).$ If the initial Dirac operator $ \dd = \ddd=
i\hbox{$ \delta ^{\mu}$}_a\gamma^a
\,{\pa}/{\pa x^{ \mu}}\,$ is the flat one, say on the (commutative) torus, then the
fluctuated one  can be any Dirac operator with arbitrary curvature and torsion,
$\dd_f=\ddd_f = i\hbox{$ e^{-1\,\mu}$}_a\gamma^a [
\,{\pa}/{\pa x^{ \mu}}\,+s(\omega _{\mu})]$. We denote by $s$ the infinitesimal
version of $S$, $e$ are the components of an orthonormal frame and $\omega $ is a
spin connection. The local group homomorphism $L$ is unique
infinitesimally \cite{uniq} and it extends the double--valued lift
$SO(3)\rightarrow SU(2)$ for Pauli spinors. Note that this last double--valuedness is
verified experimentally in neutron interferometry \cite{neu}

The definition of the spectral action
$S_\Lambda: \ff\rightarrow \rr_+$ is motivated from today's definition of the unit
of proper time in terms of the frequency of a particular atomic spectrum: 
$S_\Lambda [\dd_f]$ is the number of eigenvalues $\lambda $ of $\dd_f$ counted
with their multiplicities such that $|\lambda|\le
\Lambda $.  To compute the asymptotic behaviour of the spectral action for large
eigenvalues,
$\Lambda \rightarrow \infty$, we introduce a regulator, a differentiable function
$h:\rr_+\rightarrow\rr_+$ of sufficiently fast decrease and put 
\bb S_\Lambda ^h[\dd_f]:=\ttt\, [h(\dd^2_f/\Lambda ^2)].\ee
 If we allowed discontinuous regulators, the characteristic function of the unit 
interval for $h$ would reproduce $S_\Lambda $. With differentiable regulators, we
obtain for 4-dimensional Riemannian manifolds $M$ without torsion the following
asymptotic expansion:
\bb S_\Lambda =
\int_M [{\frac{1}{16\pi G}}\,(2\Lambda_c-R) +a(5\,R^2-8\,{\rm Ricci}^2-7\,{\rm
Riemann}^2)]\,\de V + O(\Lambda^{-2}) ,\label{heat}\ee where the cosmological
constant is $\Lambda_c= ({6h_0}/{h_2})\Lambda^2$, Newton's constant is
$G=({{3\pi }/{h_2}})\Lambda^{-2}$ and
$a={{h_4}/({5760\pi^2})}$. 
 The spectral action is universal in the sense that the regulator $h$ only enters
through its first three `moments', $h_0:=\int_0^\infty uh(u)\de u$,
$h_2:=\int_0^\infty h(u)\de u$ and $h_4=h(0)$. For small curvature, the spectral
action therefore reproduces the Einstein--Hilbert action with a positive cosmological
constant. In Robertson--Walker timespaces the curvature square term in the
spectral action vanishes identically, because up to the Euler characteristic, it is
proportional to the square of the Weyl tensor $C_{\mu \nu \rho \sigma }$. 

Now let us look at the fluctuating Dirac operators for 4-dimensional almost
commutative spaces and their spectral actions. We will find in addition to the above
gravity action an entire Yang--Mills--Higgs action including the Higgs potential
that produces spontaneous symmetry breaking.

Our first task is to lift the automorphisms of the algebra $\aaa$ in the 0-dimensional
triple to the Hilbert space. The most general such algebra is a direct sum of matrix
algebras with real, complex or quaternionic entries, $\aaa=M_n(\rr)$, $M_n(\cc),$
or $M_n(\hhh)$. All their automorphisms
$g$ in the connected component of the identity are inner, for all $a\in\aaa,$
 $g(a)=uau^{-1}$ for some unitary $u\in U(\aaa)$. This unitary is however
ambiguous by central unitaries and we have for the connected component of the
identity in the automorphism group  Aut$(\aaa)^e= O(n)/\zz_2$,
$U(n)/U(1)=SU(n)/\zz_n$ or
$USp(n)/\zz_2$.  Immediately, the following lift comes to mind,
$ L(g)=\rho (g)J\rho (g) J^{-1} $. It is at most double--valued for real and
quaternionic entries but it can have a continuous infinity of values for complex
entries. Therefore we centrally extend it \cite{fare}. Let us illustrate the central
extension for the noncommutative 0-dimensional example
Eqs.~(\ref{ncex6}--\ref{ncex11}), $\aaa=\hhh\op\cc\op M_3(\cc)$. We have 
\bb {\rm Aut}(\aaa)^e={\rm In}(\aaa)=SU(2)/\zz_2\,\times\, U(3)/U(1)
\leftarrow SU(2)/\zz_2\,\times U(3)\,\owns (v,w),\ee
 where the central unitaries in
$U(1)$ are parameterized by det$w$ and we consider the following central extensions
of the lift,
\bb L(v,w):=\rho (v,(\det w)^{q_1}, (\det w)^{q_2}\,w)\,J
\rho (v,(\det w)^{q_1}, (\det w)^{q_2}\,w)J^{-1},\label{inli}\ee with integer or
rational exponents $q_1,q_2$. For concreteness we put $q_1=1,$ $ q_2=0$.

Now let us fluctuate the Dirac operator $\dd_t=\ddd\ot 1\, +\,\gamma _5\ot\dd$
 with both timespace diffeomorphisms $\sigma $ and ``internal'' automorphisms $g$
where
$\ddd$ is again the flat Dirac operator on the torus. Note that in almost commutative
triples the internal automorphisms become timespace dependent after tensorisation,
$g$ is a gauge transformation. After some computation, we obtain the fluctuated Dirac
operator of the form  $\dd_{tf}=\ddd_f\ot 1\, +\,\gamma _5\ot\dd_f$ with
\bb \ddd_f = i\hbox{$ e^{-1\,\mu}$}_a\gamma^a [
\,{\pa}/{\pa x^{ \mu}}\,+\,s(\omega _{\mu})\,+\,\ell(A_\mu )],\qq A_\mu (x)\in
su(2)\times u(1)\times su(3).\ee We denote by $\ell$ the infinitesimal version of the
group homomorphism $L$ in Eq.~(\ref{inli}). The timespace Dirac operator becomes
covariant with respect to a Yang--Mills field $A$, that is a 1-form on the
4-dimensional commutative triple. Likewise the internal Dirac operator becomes
covariant with respect to a
$SU(2)$--doublet of complex scalar fields, that is a 1-form on the 0-dimensional
noncommutative triple:
\bb \mm_f=\pp{
\pp{\varphi _1m_u&-\bar\varphi _2 m_d\cr \varphi _2 m_u&{}\qq\bar\varphi _1
m_d}\ot 1_3 &0\cr 0&\pp{-\bar\varphi _2m_e\cr{}\qq\bar \varphi _1 m_e}}.\ee
Besides their
interpretation as a parallel transport in the discrete fifth direction,   the scalar fields
have a second interpretation: Before the fluctuation we have two parallel
4-dimensional universes a constant distance apart. After the fluctuation, this
distance becomes variable, given by the scalar fields \cite{mw}.

The configuration space of almost commutative triples is parameterized by the
gravitational degrees of freedom, a Riemannian metric given by an orthonormal
frame $e$ and a spin connection $\omega $, and by a Yang--Mills connection $A$, a
1-form valued in the Lie algebra of the automorphism group of the matrix algebra
properly lifted to the spinors and a scalar field $\varphi $ valued in a unitary
representation of this same group. This group, as well as its unitary representation
for $\varphi $, depends on the details of the chosen 0-dimensional triple.

Let us now proceed and compute the regulated spectral action asymptotically. For our
example, with vanishing torsion and $q_2=(q_1-1)/3$, we obtain:
\bb S_\Lambda ^h[\dd_{tf}]&=&\int_M[ {\frac{1}{16\pi G}}\,(2\Lambda_c-R)\, -\,a\,
C_{\mu\nu\rho\sigma}C^{\mu\nu\rho\sigma}\nonumber\\[1mm]  &&\qq
+\,1/({2}g_2^{2})\, \ttt F_{\mu\nu}^{(2)*}F^{(2)\mu\nu}+ 1/({4}g_1^{2})\, 
F_{\mu\nu}^{(1)*}F^{(1)\mu\nu}+ {1}/({2}g_3^{2})\, \ttt
F_{\mu\nu}^{(3)*}F^{(3)\mu\nu}\nonumber\\[1mm] &&\qq
+\,{\textstyle\frac{1}{2}}\,(\dee_\mu\varphi)^*
\dee^\mu\varphi\,+\,\lambda |\varphi|^4
\,-\,{\textstyle\frac{1}{2}}\,\mu^2|\varphi|^2
\,+\, {\textstyle\frac{1}{12}}\,|\varphi|^2 R\, ]\,\de V\ +\
O(\Lambda^{-2}).\label{action}\ee   
 We have decomposed the Yang--Mills
connection as $A=(A^{(2)},A^{(1)},A^{(3)})\in su(2)\times u(1)\times su(3)$, defined
the field strength $F=\de A+{\textstyle\frac{1}{2}} [A,A]={\textstyle\frac{1}{2}}
F_{\mu \nu }\de x^\mu \de x^\nu $ and the covariant derivative $\dee\varphi=\de
\varphi +A^{(2)}\varphi=
\dee_\mu \varphi \,\de x^\mu  $ with $\varphi =(\varphi _1,\varphi _2)^T$. The
coupling constants are given by:
\bb \Lambda_c=
\frac{6h_0}{h_2}\,\Lambda^2,& G={\frac{\pi }{5 h_2}}\,\Lambda^{-2},&
a={\frac{3h_4}{64\pi^2}}\,,\\[2mm] g_2^{-2}={\frac{h_4}{3\pi^2}}\,,\qq&
g_1^{-2}={\frac{5}{3}}\, {\frac{h_4}{3\pi^2}}\, q_1^2\,,\qq&
g_3^{-2}={\frac{h_4}{3\pi^2}}\,,\ee \\[-12mm]
\bb 
\lambda^{-1}={\frac{h_4}{\pi^2}}
\,\frac{(3m_u^2+3m_d^2+m_e^2)^2} {3m_u^4+3m_d^4+m_e^4}\,,\qq
\mu^2&=& 2\,\frac{h_2}{h_4}\,\Lambda^2.\ee We find indeed the entire
Yang--Mills--Higgs action as a fluctuation of the gravitational one. Of course we will
add to it the Dirac action, $S[\psi _t ,\dd_{tf}]=(\psi _t,\dd_{tf}\psi _t)$, which
contains the couplings of the graviton, of the Yang--Mills bosons  and of the Higgs
scalar to the fermions (the gravity, gauge and Yukawa couplings).

For the chosen example of the 0-dimensional, noncommutative triple and a central
 extension satisfying
$q_2=(q_1-1)/3$, we obtain the action of the standard model of electromagnetic,
weak  and strong forces with one generation of quarks and leptons and a massless
neutrino. The addition of more generations and Dirac masses in one or more
generations is straight--forward. Note however that Poincar\'e duality imposes a
purely left--handed and therefore massless neutrino in at least one generation.

So far the spectral action has been computed for only one truly
noncommutative triple, the Moyal plane, where it
produces the Moyal deformation of the Yang--Mills
action \cite{gayral}, whose quantum field theory is under active
 investigation {\it e.g.} \cite{gw}.

\section{Spontaneous symmetry breaking}

By its very construction, the spectral action is invariant under all lifted
automorphisms of its algebra. These are general coordinate transformations of
commutative timespace plus gauge transformations for almost commutative spaces.
Unlike the Euclidean Einstein--Hilbert action, the spectral action is positive and
therefore admits ground states. They are not necessarily invariant under all lifted
automorphisms, we call the stability subgroup of a given ground
state
the little group. When the little group is a proper subgroup one talks about
``spontaneous symmetry breaking from the big group down to the little group''.
Spontaneous symmetry breaking is a key ingredient of today's particle theory
because it allows to give masses to Yang--Mills bosons and to fermions, masses that
are ruled out by gauge invariance. In the following we simplify the configuration
space by deleting gravity: we take timespace to be the flat 4-dimensional torus of
fixed volume and look for the minima of the remainder of the spectral action
(\ref{action}). Putting the connection $A$ to zero and the scalar field $\varphi $ to
be constant minimizes the action if the norm satisfies
\bb |\stackrel{\circ}{\varphi} |^2=|\stackrel{\circ}{\varphi}
_1|^2+|\stackrel{\circ}{\varphi} _2|^2=\mu ^2/(4\lambda ).\ee 
A constant scalar field
$\stackrel{\circ}{\varphi} $ with this norm is called vacuum expectation value. It
breaks the gauge invariance spontaneously with little group
$U(3)$ in our example. Expanding the scalar field around such a minimum,
$\varphi(x) =\stackrel{\circ}{\varphi} +H(x)$ then induces masses for the
Yang--Mills bosons outside the little group via the term
${\textstyle\frac{1}{2}}\,(\dee_\mu\stackrel{\circ}{\varphi})^*
\dee^\mu\stackrel{\circ}{\varphi}$, masses for the fermions via the Yukawa
couplings $(\psi ,\stackrel{\circ}{\dd_f}\psi )$ and masses to the Higgs
scalar $H$ via the Higgs potential $\lambda |\varphi |^4-{\textstyle\frac{1}{2}} \mu
^2|\varphi |^2.$ The fermion masses are (the absolute values of) the eigenvalues of
the fluctuated minimal internal Dirac operator
$\stackrel{\circ}{\dd_f}$. Therefore the fermion masses are not given  by
the parameters
$m_u,\ m_d,\ m_e$ in the initial fermionic mass matrix $\mm$, Eq.~(\ref{mass}), but
from the parameters  in the minimum of the fluctuated one,
$\stackrel{\circ}{\mm_f}$. In the standard model, they happen to be
identical. This coincidence is far from being generic. In the generic case, you will
start with different masses in one multiplet, say
$m_u<m_d$ because this is what experiment tells you. But after fluctuating and
minimizing you will get degenerate masses, $\stackrel{\circ}{
m_u}=\stackrel{\circ}{ m_d}$. This mass degeneracy is in conflict with
quantum corrections if the subgroup defining the multiplet, here $SU(2)$, is
spontaneously broken. The initial and all fluctuated internal Dirac operators have
degenerate eigenvalues. Nonvanishing masses appear four times, vanishing ones
twice. In the standard model, Eq.~(\ref{mass}), we have an additional threefold
degeneracy of the the quark masses $m_u$ and $m_d$ because of the $\ot 1_3$. ``
Quarks are $SU(3)$ (colour) triplets''. All these degeneracies follow from the axioms
of spectral triples: $\dd$ commutes with the real structure, anticommutes with
chirality and the first order axiom implies the colour degeneracy. Therefore we call
these kinematical degeneracies. Generically there will be additional `dynamical'
degeneracies following from the minimisation of the action, not so in the standard
model with one generation.

The standard model has two other remarkable properties, that we will discuss now.

\section{A natural selection of the standard model}

A real, even spectral triple is called {\bf $S^0$-real} \cite{real} if there is a unitary
operator
$\epsilon $ on $\hh$, called $S^0$-real structure, satisfying 
\bb \epsilon ^2=1,\qq [\epsilon ,\rho (a)]=0 \
{\rm for\ all}\ a\in\aaa,\qq [\epsilon ,\dd]=[\epsilon ,\chi ]=0,\qq
\epsilon J=-J\epsilon.\ee 
The commutative triples of Riemannian spaces are never $S^0$-real. For
0-dimensional triples the $S^0$-reality has a physical interpretation, it excludes the
existence of Majorana--Weyl fermions, which are allowed in 0 mod 8 dimensional
Euclidean space or more generally in Minkowski spaces with a difference of plus
and minus signs equal to 0 mod 8.

The 0-dimensional triple of the standard model is $S^0$-real,
\bb\epsilon =\pp{1_{15}&0\cr 0&-1_{15}}.\ee

We call a spectral triple $(\aaa,\hh,\dd)$  {\bf reducible} if
there is a proper subspace
$\hh_0\subset\hh$ invariant under the algebra $\rho(\aaa)$  such that
$(\aaa,\hh_0,\dd|_{\hh_0})$ is a spectral triple. If the
triple is real, $S^0$-real and even, we require  the subspace
$\hh_0$ to be also invariant under the real structure $J$, the $ S^0$-real
structure $\epsilon $ and under the chirality
$\chi $ such that the triple $(\aaa,\hh_0,\dd|_{\hh_0})$ is again real,
$S^0$-real and even. 

The commutative triples of Riemannian spaces are all irreducible. The triple of the
standard model with one generation of quarks and leptons is also irreducible.

Let us summarize, the standard model with one generation of fermions has three
properties, it is
$ S^0$-real, irreducible and dynamically non--degenerate. It is natural to ask
whether there are many other such triples. Here is a partial answer \cite{class}: 

The sum of matrix algebras,
$\aaa=\bigoplus_{i=1}^N \aaa_i$ with $N=1,2,3$ admits a 0-dimensional, real,
$S^0$-real,
 irreducible and dynamically non--degenerate spectral triple only if
it is in the list
\begin{center}
\begin{tabular}{|c|c|c|}
\hline &&\\[-2mm]
 $N=1$&$N=2$&$N=3$\\[1ex]
\hline &&\\[-2mm]
 &&${\bf 1}\op{\bf 1}\op\ccc$\\[1ex]
 {\rm void}&${\bf 2}\op{\bf 1}$&${\bf 2}\op{\bf 1}\op\ccc$\\[1ex]
 &&${\bf 2}\op{\bf 1}\op {\bf 1}$\\[1ex]
\hline
\end{tabular}
\end{center} Here $\bf 1$ is a short hand for $\rr$ or $\cc$ and $\bf 2$
for $M_2(\rr)$,
$M_2(\cc)$ or
$\hhh$.\\ The `colour' algebra
$\ccc$ is any simple matrix algebra and has two important constraints:\\ 
 i) Its
representation is ``vector--like'', identical on corresponding left- and right-handed
subspaces of
$\hh$. \\
 ii) The
Dirac operator $\dd $  is invariant under $U(\ccc)$,
$\rho (1,1,w) \, \dd \, \rho (1,1,w)^{-1}=\dd,$ for all $w\in
U(\ccc)$. This implies that the unitaries of $\ccc$ do not participate
in the fluctuations and are therefore unbroken, i.e. elements of the little
group.

The triple with two simple algebras is the two--point space where however one
point has been rendered noncommutative in order to have algebra
automorphisms in the connected component of the identity.

Let us conclude this section with the Darwinean explanation of gravity: In the
beginning apples fell in all directions, only those falling towards the earth
survived a natural selection.

\end{document}